\documentclass[twocolumn,superscriptaddress,times]{revtex4-1}
\usepackage{graphics,graphicx,epsfig,ulem,epstopdf,bm,longtable,url,geometry,datetime,physics}
\usepackage[dvipdfm,colorlinks=true,linkcolor=blue,urlcolor=blue,citecolor=blue]{hyperref}
\usepackage{amsmath,amssymb,latexsym,float,amsfonts,subfigure,hyperref,color,setspace}
\usepackage[ansinew]{inputenc}
\usepackage[usenames,dvipsnames]{pstricks}
\usepackage[pagewise,mathlines]{lineno}

\def\footnoterule{\kern -1mm \hrule width 6.0cm \kern 2.2mm}%
\usepackage{tikz,xcolor,hyperref}
\definecolor{lime}{HTML}{A6CE39}
\DeclareRobustCommand{\orcidicon}{%
    \begin{tikzpicture}
    \draw[lime, fill=lime] (0,0)
    circle [radius=0.16]
    node[white] {{\fontfamily{qag}\selectfont \tiny ID}};\draw[white, fill=white] (-0.0625,0.095)
    circle [radius=0.007];
    \end{tikzpicture}
    \hspace{-2mm}}
\foreach \x in {A, ..., Z}
{\expandafter\xdef\csname orcid\x\endcsname{\noexpand\href{https://orcid.org/\csname orcidauthor\x\endcsname}{\noexpand\orcidicon}}}

\begin{document}


\title{\bf\fontsize{12}{12} Rapid and Stable Collective Charging and Discharge Suppression in Strongly Coupled Many-Body Quantum Batteries}

\author{Shun-Cai Zhao\orcidA{}}
\email[Corresponding author: ]{zsczhao@126.com}
\affiliation{Center for Quantum Materials and Computational Condensed Matter Physics, Faculty of Science, Kunming University of Science and Technology, Kunming, 650500, PR China}

\author{Yi-Fan Yang}\email[Co-first author.]{}
\affiliation{Center for Quantum Materials and Computational Condensed Matter Physics, Faculty of Science, Kunming University of Science and Technology, Kunming, 650500, PR China}

\author{Ni-Ya Zhuang}
\affiliation{Center for Quantum Materials and Computational Condensed Matter Physics, Faculty of Science, Kunming University of Science and Technology, Kunming, 650500, PR China}


\begin{abstract}
Achieving rapid and stable energy storage in quantum batteries (QBs) remains a key challenge, particularly under strong system-environment coupling where non-Markovian effects become prominent. While most previous studies focus on weak coupling regimes, we propose a many-body QB model exhibiting collective charging and discharge suppression in a non-perturbative regime. The model adopts a $\Lambda$-type configuration where multiple battery units share a common excited state and have individual ground states, forming an effective collective structure. To accurately capture the dynamics under strong coupling, the system's time evolution is governed by a Redfield-type master equation incorporating memory effects via a Debye spectral density. We quantify the stored energy using ergotropy and analyze the impact of tunneling, driving strength, spectral width, and environmental temperature on charging performance. Numerical simulations reveal that optimized driving and reservoir engineering can simultaneously achieve rapid and stable charging while suppressing energy leakage. These results provide theoretical insight into strong-coupling thermodynamics and guide the design of robust QB platforms using solid-state or atomic systems.
\begin{description}
\item[Keywords] Many-body quantum battery; ergotropy dynamics; rapid-stable charging; \\  strong coupling; non-Markovian dynamics.
\end{description}
\end{abstract}

\date{\today}
\maketitle

\section{Introduction}\label{Introduction}

Quantum batteries (QBs), envisioned as atomic-scale energy storage and conversion devices, have emerged as a promising component in quantum technologies~\cite{Alicki2012EntanglementBF,Andolina2018ExtractableWT}. The miniaturization of quantum devices~\cite{Xu2020SecureQK} has motivated fundamental investigations into whether quantum coherence and correlations can be exploited to enhance charging efficiency~\cite{Rossini2020QuantumAI,Santos2021QuantumAO}. Two principal charging paradigms have been proposed: the \textit{parallel charging} scheme, where each quantum cell is charged independently~\cite{Santos2021QuantumAO}, and the \textit{collective charging} scheme, where a global unitary drives the entire system coherently~\cite{Rossini2020QuantumCS,Gyhm2021QuantumCA}. Recent efforts have revealed that collective protocols can yield a quantum advantage in both charging speed and power~\cite{Campaioli2018QuantumB,Farina2019ChargerM,Rossini2019ManyBodyL,JuliaFarre2020BoundsO}.

In addition to charging performance, researchers have explored energy extraction from QBs in realistic noisy environments. Tirone et al.~\cite{Tirone2023WorkExtraction} showed that nonlocal correlations can mitigate noise-induced degradation during work extraction. Similarly, remote charging schemes~\cite{Song2023RemoteCharging} and coherent collision models~\cite{Seah2021QuantumSpeedUp} have been introduced to suppress energy loss and improve the robustness of quantum batteries. In particular, the role of quantum coherence in enabling faster charging and higher ergotropy has been highlighted~\cite{Cakmak2020Ergotropy}.

Despite these advances, most previous studies assume weak system-environment coupling and Markovian dynamics~\cite{Ju2018Bounds,Le2017SpinChain,Zhang2018PowerfulHarmonic}. However, in practical implementations-such as quantum dots, superconducting circuits, or Rydberg atom arrays-quantum systems often interact strongly with their environments, resulting in non-Markovian dynamics. The validity of conventional Lindblad or Born-Markov master equations becomes questionable in such regimes~\cite{Andolina2018QuantumVsClassical,Kamin2020Entanglement}. Thus, a key open question is: can collective effects still offer performance advantages under strong coupling and memory effects ? And if so, how can energy leakage or reverse flows be effectively suppressed ?

In this work, we propose a strongly coupled many-body QB model operating in a non-perturbative regime. The model adopts a $\Lambda$-type collective configuration, in which multiple QBs share a common excited state while maintaining distinct ground states. The system's dynamics is governed by a Redfield-type master equation incorporating non-Markovian effects through spectral engineering. We quantify the stored energy and extractable work (ergotropy) and systematically analyze how various physical parameters-such as driving strength, energy gap, tunneling interaction, and environment temperature-affect the charging performance.

The rest of this paper is organized as follows: In Sec.~\ref{sec:model}, we construct a many-body collective QB model with effective two-level structure. In Sec.~\ref{sec:charging_dynamics} and \ref{sec:three-qubit QBs}, the energy storage and work extraction dynamics are formulated based on ergotropy. In Sec.~\ref{sec:Results}, we analyze a three-qubit QB example to study rapid-stable charging and discharge suppression. Finally, we summarize our findings in Sec.~\ref{sec:conclusion}.

\section{Many-body Collective QB Model}\label{sec:model}

\begin{figure}[htbp]
\center
\centering
\includegraphics[width=0.9\columnwidth] {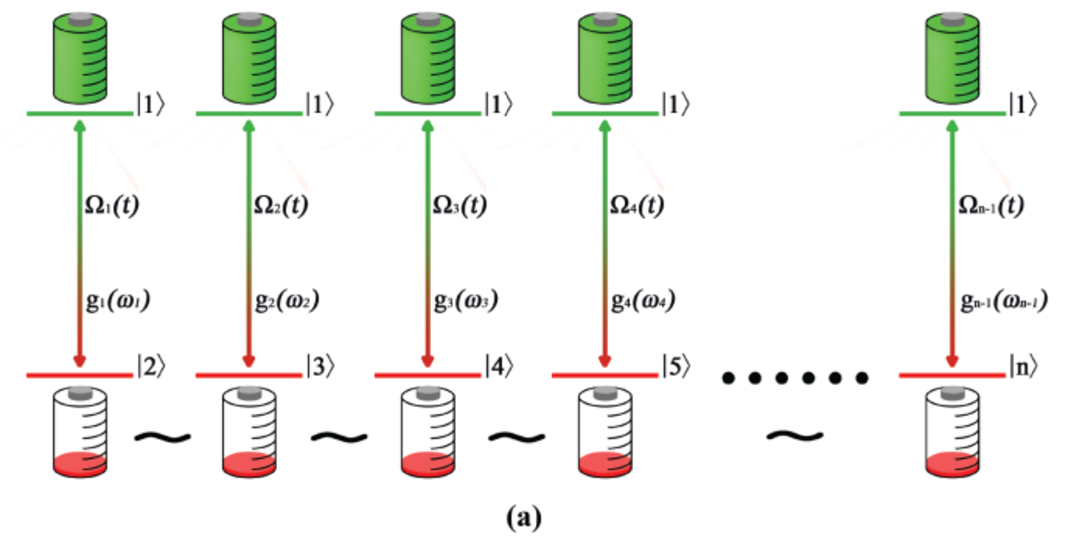}\hspace{0.5cm}\includegraphics[width=0.9\columnwidth] {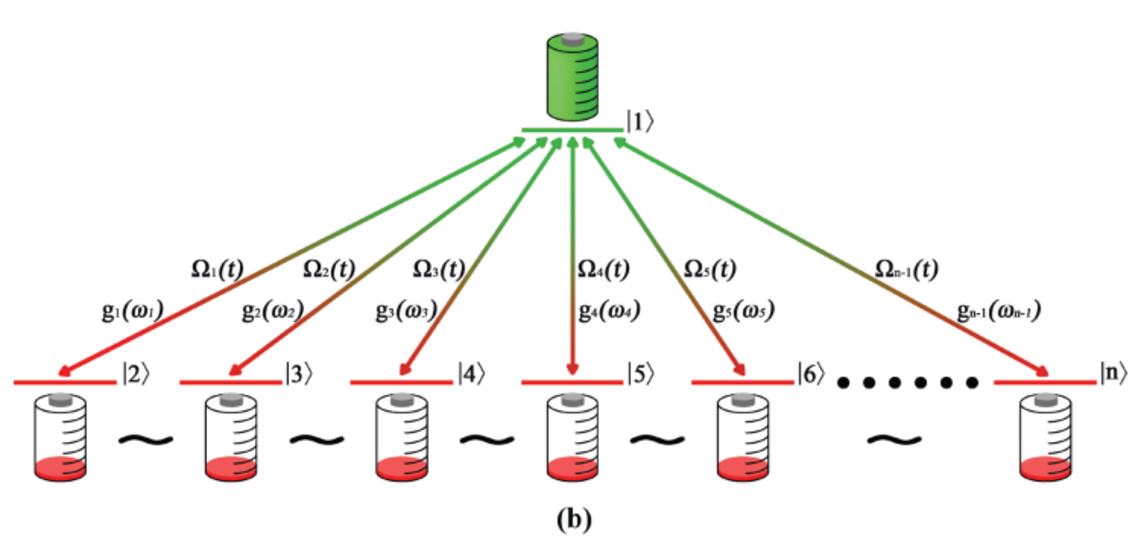}\hspace{0in}%
\caption{Parallel (top) versus collective (bottom) charging schemes. (a) Multiple QBs represented by TLSs are placed side by side for parallel charging with quantum tunneling effects between each battery, (b) the simplified collective charging scheme. Where $\Omega_i(t)$ is the charging function for the external field charging the $i$h QB. $g_{i}(\omega_i)$ is the coupling function for the interaction between the $i$-th QB and the external field, and the wavy lines between adjacent QBs represent the quantum tunneling effects existing between the two QBs.}\label{Fig1}
\end{figure}

To explore QB behavior beyond weak coupling, we consider a one-dimensional many-body QB system strongly coupled to a bosonic environment. Each unit cell (battery) features a common excited state $\ket{1}$ and an individual ground state $\ket{i}$ $(i=2,\dots,n)$, forming an effective $\Lambda$-type structure [see Fig.~\ref{Fig1}]. The level $\ket{1}$ represents the fully charged state, while the ground states $\ket{i}$ $(i = 2, 3, \dots, n)$ correspond to the fully discharged configurations, as shown in the model of Fig.~\ref{Fig1}(a). We denote the time-dependent charging field applied to each battery unit by $\Omega_i(t)$ for $i = 2, 3, \dots, n{-}1$. The system-environment coupling strength for each unit is described by $g_i(\omega_i)$, while the tunneling interaction between adjacent battery units is uniformly characterized by a constant coupling parameter $T_e$. This abstraction captures the essence of collective charging dynamics, where multiple units interact coherently with shared degrees of freedom. Similar configurations can be realized in quantum dots~\cite{PhysRevA.107.023725}, Rydberg atoms~\cite{mondal2021manybody}, or trapped ions with symmetric couplings.

The total Hamiltonian is given by:
\begin{equation}
    \hat{H} = \hat{H}_\text{S} + \hat{H}_\text{E} + \hat{H}_\text{int},  \label{eq1}
\end{equation}
where $\hat{H}_\text{S}$ describes the QB system, $\hat{H}_\text{E}$ the environment, and $\hat{H}_\text{int}$ the interaction between them.

The system Hamiltonian incorporates time-dependent driving and tunneling between neighboring battery units:
\begin{equation}
    \hat{H}_\text{S} = \sum_{i=1}^{n} \varepsilon_i \hat{\sigma}_i + \sum_{j=2}^{n} \Omega_j(t) \ket{1}\bra{j} + \sum_{j=2}^{n-1} T_\text{e} \ket{j}\bra{j+1} + \text{H.c.},   \label{eq2}
\end{equation}
where $\Omega_j(t)$ denotes the external driving field that governs the charging process. The driving function $\Omega_j(t)$ is typically sinusoidal or periodic to mimic realistic modulation. $T_\text{e}$ represents quantum tunneling, and $\varepsilon_i$ the level energies.

The environment is modeled as a bath of harmonic oscillators, consistent with the Debye model~\cite{kubo1966fluctuation}:
\begin{equation}
    \hat{H}_\text{E} = \sum_{k} \omega_k \hat{a}_k^\dagger \hat{a}_k,                \label{eq3}
\end{equation}
with cutoff frequency $\omega_D$, typically satisfying \(\omega_k \leq \omega_D \). The system-environment interaction is:
\begin{equation}
    \hat{H}_\text{int} = \sum_k g_k \hat{A}_\text{S} \otimes \hat{B}_k,               \label{eq4}
\end{equation}
where $\hat{A}_\text{S}$ is a system operator (e.g., $\hat{\sigma}_z$ or projection operators) and $\hat{B}_k$ denotes bath degrees of freedom (e.g., displacement or number operators).
 Here, \( g_{k}\) represents the coupling strength between the system and the environment, which typically depends on the wave vector \( \mathbf{k} \). \( \hat{A}_S \) is an operator in the QBs, representing its coupling with the environment, while the degrees of freedom of the environment are described by \(\hat{B}_k \). In the Debye model, the environment is associated with phonon-related operators, such as the displacement operator of the \( k \)-th mode in the environment.

To capture open-system dynamics under strong coupling, we adopt a Redfield-type master equation\cite{Crowder2023InvalidationOT,Becker2022CanonicallyCQ,Zhao2016Dynamics}:
\begin{eqnarray}
\frac{d}{dt} \hat{\rho}(t)=& -\frac{i}{\hbar} \left[\hat{H}_\text{S}(t), \hat{\rho}(t) \right] + \sum_{i,j} R_{ij} ( 2 \hat{L}_{ij} \hat{\rho}(t) \hat{L}_{ij}^\dagger \label{eq5}\\
                           &- \left\{ \hat{L}_{ij}^\dagger \hat{L}_{ij}, \hat{\rho}(t) \right\} ),\nonumber
\end{eqnarray}

\noindent where $\hat{L}{ij} = \ket{i}\bra{j}$ denote Lindblad-like jump operators, and $R_{ij}(\omega)$ is the Redfield tensor element that characterizes dissipative transitions between states. In contrast to Markovian Lindblad forms, the Redfield tensor explicitly depends on both the environmental spectral density and temperature:
\begin{equation}
R_{ij}(\omega) = J_{ij}(\omega) \left[ \coth\left( \frac{\hbar \omega}{2 k_B T} \right) + 1 \right],       \label{eq6}
\end{equation}
where $J_{ij}(\omega)$ is the spectral density describing the coupling between the system and the environment. In this work, we consider a Debye-type form:
\begin{equation}
J_{ij}(\omega) = \gamma_{ij} \frac{\omega}{\omega_0^2 + \omega^2},                                          \label{eq7}
\end{equation}
which captures the frequency-dependent interaction strength between the system and its environment, and reflects the finite memory (correlation) time of the bath. Spectral density functions such as the Debye~\cite{Kubo1966TheFT}, Ohmic~\cite{Leggett1987DynamicsOT}, and Drude-Lorentz~\cite{Drude1900ZurET} models are widely employed in modeling different types of system-bath couplings in quantum open systems. The specific form of $J_{ij}(\omega)$  directly determines the structure of the Redfield tensor and thereby governs the dissipative dynamics of the system.

\begin{figure}[htbp]
\centering
\includegraphics[width=0.8\columnwidth] {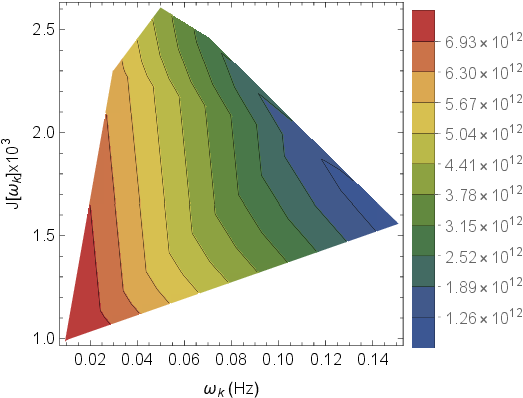}\hspace{0in}%
\caption{Redfield tensor vs the environmental frequency \(\omega_{k}\) and Debye environmental spectral density \(J (\omega_{k})\) with other parameters being \(\gamma=2.6\times10^{-4}Hz, T=300K,\omega_{0}=0.05 Hz\).}\label{Fig2}
\end{figure}

To illustrate this effect, Fig.~\ref{Fig2} shows the behavior of the Redfield tensor $R_{ij}(\omega)$   at room temperature, evaluated using the Debye spectral density. As depicted,
$J_{ij}(\omega)$ remains finite over the relevant frequency range, and both the environmental frequency \(\omega_{k}\) and the spectral profile contribute to a suppression of the dissipative coefficients. This demonstrates how environmental structure modulates the decoherence and energy relaxation in the quantum battery system.

This Redfield-based framework enables the incorporation of finite bath correlation times and partially captures non-Markovian memory effects and moderate-to-strong system-environment coupling. Although the Redfield master equation is perturbative in nature, it has been successfully applied to regimes beyond strict weak coupling, especially in the presence of structured or frequency-selective environments. This motivates its adoption here as a compromise between tractability and physical realism. While it does not fully account for all non-perturbative features, recent studies have demonstrated its applicability in structured environments where traditional Born-Markov approximations break down~~\cite{Crowder2023InvalidationOT,Becker2022CanonicallyCQ}. In this work, we restrict our analysis to regimes where the Redfield equation remains a valid approximation, as supported by these references. The master equation is numerically integrated using time-discretized evolution with physically motivated initial conditions and enforced trace preservation.

\section{Energy Quantification}\label{sec:charging_dynamics}

To evaluate the performance of a QB, it is essential to monitor both the internal energy storage and the dissipative losses to the environment throughout its evolution. In realistic scenarios involving strong system-environment coupling, such an evaluation requires a theoretical treatment that goes beyond standard weak-coupling or Markovian approximations, as non-Markovian memory effects become significant. To address this, we employ a Redfield-type quantum master equation (QME) [Eq.~\eqref{eq5}], which incorporates frequency-dependent dissipation through structured spectral densities. This approach is supported by recent studies targeting intermediate-to-strong coupling regimes~\cite{Crowder2023InvalidationOT,Becker2022CanonicallyCQ}.

We quantify the QB's performance using two key observables: internal energy and ergotropy.

\paragraph*{Internal energy.} The energy stored in the QB at time $t$ is given by:
\begin{equation}
E(t) = \text{Tr}\left[ \hat{H}_\text{S} \hat{\rho}(t) \right] - \text{Tr}\left[ \hat{H}_\text{S} \hat{\rho}(0) \right], \label{eq8}
\end{equation}
which measures the net energy gained relative to the initial state.

\paragraph*{Ergotropy.} To evaluate the extractable work under unitary operations, we employ the notion of 	extit {ergotropy}~\cite{allahverdyan2004maximal}, defined as the energy difference between the system¡¯s instantaneous state and its passive counterpart:
\begin{equation}
\mathcal{E}(t) = \text{Tr}\left[ \hat{H}_\text{S} \hat{\rho}(t) \right] - \text{Tr}\left[ \hat{H}_\text{S} \hat{\rho}_\text{pass}(t) \right], \label{eq9}
\end{equation}
where $\hat{\rho}_\text{pass}(t)$ denotes the passive state associated with $\hat{\rho}(t)$. This quantity represents the portion of energy that can be converted into useful work, excluding thermal or disordered contributions.

All numerical simulations are performed via time-discretized integration of the Redfield-type QME [Eq.~\eqref{eq5}], ensuring trace preservation and physical consistency of $\hat{\rho}(t)$ throughout the evolution.

Regarding the system configuration, we consider a $\Lambda$-type level structure where each QB possesses a distinct ground state $\ket{i}$ and shares a common excited state $\ket{1}$. Such configurations can be physically realized in multi-level quantum dot arrays~\cite{PhysRevA.107.023725}, engineered spin-1 systems~\cite{shi2022entanglement}, or Rydberg atom platforms~\cite{mondal2021manybody}. Even when idealized, this structure captures essential features of coherence-enhanced storage and collective dynamics, and is widely used in quantum optics for modeling effects such as dark-state superpositions and electromagnetically induced transparency (EIT)~\cite{fleischhauer2005eit}.

Before proceeding to numerical analysis, we illustrate the model behavior through a minimal collective configuration. A three-qubit example is introduced to explicitly examine the dynamical features of rapid charging and energy dissipation regulation under various parameter regimes. In the next section, we will numerically analyze the time evolution of the internal energy $E(t)$ and ergotropy $\mathcal{E}(t)$ under varying control parameters, including tunneling strength $T_e$, spectral cutoff $\omega_0$, and driving amplitude $V$.

\section{Charge-discharge dynamics of collective three-qubit QBs }\label{sec:three-qubit QBs}

To measure the charging and discharging dynamics of many-body QBs, we will take the charging and discharging behavior of a three-qubit QB system (i.e., $n=4$) as an example, focusing on how many-body QBs achieve fast charging and the control strategies for counteracting mechanisms that hinder rapid charging. According to Fig.~\ref{Fig1}, we can write the Hamiltonian of the three-qubit QB system as,

\begin{eqnarray}
\hat{H}_{s} =
\begin{bmatrix}
\varepsilon_{1} & \Omega_{12}(t) & \Omega_{13}(t) & \Omega_{14}(t) \\
\Omega_{21}(t) & \varepsilon_{2} & T_{e} & 0 \\
\Omega_{31}(t) &  T_{e} & \varepsilon_{3} & T_{e} \\
\Omega_{41}(t) & 0 & T_{e} & \varepsilon_{4}
\end{bmatrix}\label{eq9}
\end{eqnarray}

\noindent In Eq.~(\ref{eq9}), the charging function in the Hamiltonian are defined as,
\begin{eqnarray}
&&\Omega_{12}(t)=\Omega_{21}(t)=V\sin(\Omega {t}/\tau)\\
&&\Omega_{13}(t)=\Omega_{31}(t)=V[1-\cos(\Omega {t}/\tau)]\\
&&\Omega_{14}(t)=\Omega_{41}(t)=V\sin( \Omega {t}/\tau)
\end{eqnarray}

\noindent where $V$ is the amplitude of charging function, $\Omega$ is an integer and $\tau$ is the maximum charging time. \(\varepsilon_{2}\) = \(\varepsilon_{3}\) = \(\varepsilon_{4}\) = \(0.25 eV \), \(\varepsilon_{1}\) = \(0.25 eV \) + \(\Delta E\), and \(\Delta E\) represents the energy gap. \(T_{e}\) is the tunneling effect between different individual battery units.

\section{Results and discussions}\label{sec:Results}

Next, we will focus on examining the dynamical behavior of energy storage in three parallel QBs. Before starting the numerical calculations, some typical parameters need to be predetermined. They are listed in detail in Tab.~\ref{Tab.1}.

\begin{table}
\begin{center}
\caption{\small Parameters for the collective three-qubit QBs.}
\label{Tab.1}
\vskip 0.02cm\setlength{\tabcolsep}{1.0pt}
\resizebox{1\linewidth}{!}{
\begin{tabular}{  c|c|c|c|c|c|c|c|c }
\hline
         & \(\Omega\)  & \(\Delta{E}(eV) \) &\(\omega (Hz)\)& \(\gamma\times10^{-7} (s) \) & \(\omega_{0} (Hz)\)& \(Te \)&\(T (k)\) & \(V (\mu V) \)\\
\hline
Fig.3(a) & $\setminus$ &        1.5         &    0.085      &$2.6 $ &     0.10        &   0    &  300     &   1.5     \\

Fig.3(b) & \(1.0\pi\)  &    $\setminus$     &    0.085      &$2.6 $ &     0.12        &   0    &  300     &   1.5     \\

Fig.3(c) & \(1.0\pi\)  &     2.75           &    0.085      &$2.6 $ &     0.12        &   0    &  300     & $\setminus$ \\

Fig.3(d) & \(1.0\pi\)  &     2.75           &    0.085      &$2.6 $ &     0.12        & $\setminus$& 300  &   1.5     \\

Fig.4(a) & \(1.0\pi\)  &     1.5            &    0.085      &$2.6 $ & $\setminus$     &   0    &  300     &   1.5    \\

Fig.4(b) & \(1.0\pi\)  &     1.5            & $\setminus$   &$9.0 $ &    0.12         &   0    &  300     &   1.5    \\

Fig.4(c) & \(1.0\pi\)  &     1.5           &     0.085      &$\setminus$ &  0.03      &   0    &  300     &   1.5    \\

Fig.4(d) & \(1.0\pi\)  &     1.5           &     0.143      &$9.0 $ &0.08             &   0    &$\setminus$&  1.5    \\
\hline
\end{tabular}}
\end{center}
\end{table}
\subsection{Dynamics of Ergotropy under Driving and Intrinsic Parameter Control}

\begin{figure}[htbp]
\centering
\begin{minipage}{0.9\columnwidth}
\centering
\includegraphics[width=0.5\columnwidth] {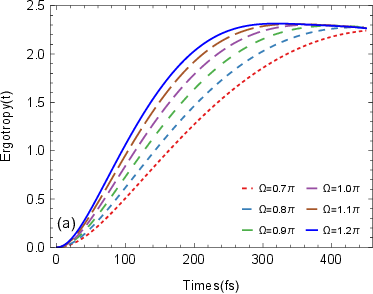}\hspace{0in}\includegraphics[width=0.50\columnwidth] {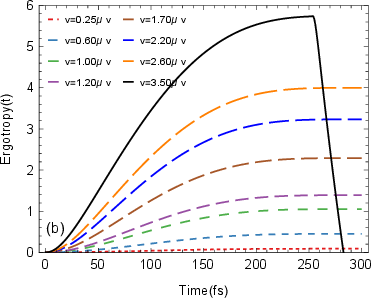}\hspace{0in}%
\end{minipage}
\hspace{-0.5cm} 
\begin{minipage}{0.9\columnwidth}
\centering
\includegraphics[width=0.5\columnwidth] {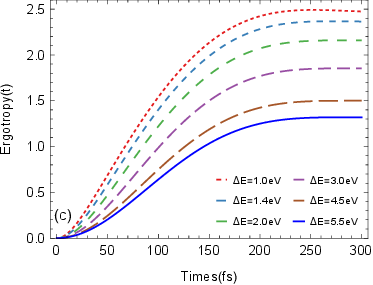}\hspace{0in}\includegraphics[width=0.50\columnwidth] {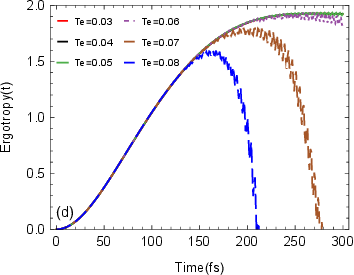}\hspace{0in}%
\caption{Time evolution of ergotropy in the collective three-qubit QB model under different control parameters: (a) frequency $\Omega$ of the charging field; (b) amplitude $V$ of the charging field; (c) bandgap $\Delta E$ of the QB material; (d) inter-QB tunneling strength $T_e$.}\label{Fig3}
\end{minipage}
\end{figure}

The capability of a QB to undergo rapid-stable charging is a crucial metric of its operational performance. In this section, we examine how various system and driving parameters-including the frequency and amplitude of the charging field, the material bandgap, and the inter-battery tunneling strength-modulate the time-dependent evolution of ergotropy. The results are presented in Fig.~\ref{Fig3}, where each subfigure corresponds to a specific control parameter.

\paragraph*{(a) Frequency Dependence.}
As shown in Fig.~\ref{Fig3}(a), increasing the driving frequency $\Omega$ significantly accelerates the time required for the QB to reach a steady ergotropy plateau. When $\Omega$ increases from $0.7\pi$ to $1.2\pi$, the system reaches its stable charging state $180$~fs earlier, corresponding to a $40\%$ reduction in total energy storage time. This acceleration indicates that higher driving frequencies enhance coherent energy transfer between the charging field and the many-body QB system. The frequency $\Omega$ thus serves as a tunable parameter to facilitate rapid-stable charging in practical implementations.

The enhanced stability observed with increasing frequency $\Omega$ in Fig.~\ref{Fig3}(a) reflects more efficient resonance coupling between the QB and the driving field. When $\Omega$ approaches the system's transition frequency, it facilitates coherent energy transfer and suppresses low-frequency fluctuations, thereby shortening the time to reach stable ergotropy. From a dynamical perspective, higher $\Omega$ corresponds to sampling higher spectral density regions in the environment, reducing entropy exchange and enhancing the coherence timescale. This behavior highlights the role of resonance engineering in achieving rapid-stable charging.

\paragraph*{(b) Amplitude Dependence.}
Fig.~\ref{Fig3}(b) shows the effect of the charging field amplitude $V$ on the ergotropy evolution. When $V$ is relatively small, the peak ergotropy grows nearly linearly with amplitude, while the time to reach peak remains approximately invariant. This behavior resembles classical expectations of proportional energy accumulation. However, a striking deviation occurs when $V$ exceeds a critical value-e.g., $V = 3.5~\mu$V-where the stable charging behavior is disrupted, and the ergotropy decays rapidly to zero. This phenomenon reflects a breakdown of the coherent energy transfer mechanism, likely due to resonance-induced energy leakage or destructive interference among system components. Such nonlinear behavior is a feature of many-body QB dynamics and has no classical analogue.

While Fig.~\ref{Fig3}(b) shows linear scaling of peak ergotropy with $V$ at small amplitudes, a critical threshold ($V$ = 3.5~$\mu$V) marks the onset of charging instability. Physically, this arises from the breakdown of the linear response regime. Large $V$ induces strong transitions across multiple energy levels, amplifying interference pathways and enabling destructive inter-qubit correlations. Moreover, excessive amplitude may lead to non-adiabatic transitions akin to the Landau-Zener process\cite{altland2008nonadiabacity}, thereby interrupting energy storage and partially reversing work extraction. This observation serves as a caution for the optimal regime of driving strength in QB operation.

\paragraph*{(c) Bandgap Effect.}
The intrinsic bandgap $\Delta E$ of the material forming the QB primarily affects the extractable energy rather than the charging time. As shown in Fig.~\ref{Fig3}(c), varying $\Delta E$ has minimal influence on the time needed to reach maximum ergotropy. However, larger bandgaps correlate with lower peak ergotropy values. This negative correlation is consistent with prior findings~\cite{Zhao2019EnhancedQY} that wider bandgaps suppress the optical transition rates, thereby limiting the system's ability to store extractable energy. This observation provides a crucial material selection criterion for the practical realization of quantum batteries: materials with moderate bandgaps may offer superior performance.

As depicted in Fig.~\ref{Fig3}(c), increasing the intrinsic bandgap $\Delta E$ decreases the peak ergotropy while leaving the charging timescale unchanged. This outcome is consistent with the fact that larger gaps inhibit thermal and optical excitation, thereby reducing extractable energy. However, since the Redfield master equation governs relaxation via system-environment coupling, the temporal dynamics remain largely unaffected. These results suggest that bandgap engineering is crucial for maximizing QB capacity, and materials with moderate $\Delta E$ may offer optimal performance.

\paragraph*{(d) Tunneling-Induced Dissipation.}
The tunneling interaction between neighboring QB units introduces quantum correlations and dissipative effects. Fig.~\ref{Fig3}(d) illustrates that for weak tunneling strengths ($T_e$=0.03,0.04,0.05), the ergotropy displays creasing oscillations, indicative of under-damped coherent energy exchange. As $T_e$ increases ($T_e$=0.06,0.07,0.08), a transition occurs: the ergotropy rapidly decays to zero within a short time, signifying a crossover to overdamped, dissipation-dominated dynamics. This critical behavior highlights the dual role of tunneling: while moderate tunneling can sustain inter-battery coherence, excessive tunneling introduces decoherence that impedes energy storage. Therefore, controlled engineering of inter-unit coupling is essential to optimize charging stability and output power in many-body QB architectures~\cite{Zhong2021PhotovoltaicPE}.

Fig.~\ref{Fig3}(d) shows that small tunneling strength $T_e$ sustains coherent energy exchange, manifested as increasing oscillations in ergotropy. As $T_e$ increases, a rapid transition occurs-ergotropy decays to zero, signifying over-damped dynamics. This coherence-to-decoherence crossover stems from the enhanced inter-battery energy leakage, whereby tunneling induces dephasing and accelerates entropy production. The observed threshold behavior underscores the dual role of tunneling: a moderate $T_e$ is beneficial for energy transfer, while excessive tunneling leads to energy dissipation and degraded QB performance. This nontrivial result reveals a tunable pathway to control coherence preservation in large-scale QB architectures.

The above analyses collectively reveal a key insight: optimal energy storage requires balancing driving-field strength, material characteristics, and inter-unit coherence. Both overdriving and over-coupling can lead to destructive quantum interference and dissipative losses. Hence, engineering QBs involves not only selecting suitable parameters but also understanding their interplay under the framework of non-Markovian quantum thermodynamics. These findings offer design principles for the development of robust and scalable QB platforms.

\subsection{Dynamics of Ergotropy under Environmental Parameters}

\begin{figure}[ht]
\centering
\begin{minipage}{0.9\columnwidth}\centering
\includegraphics[width=0.50\columnwidth] {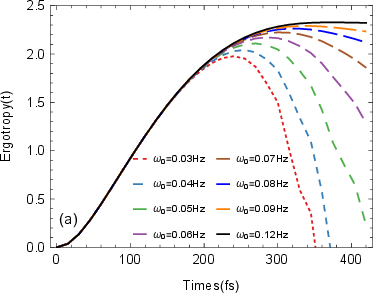}\hspace{0in}\includegraphics[width=0.50\columnwidth] {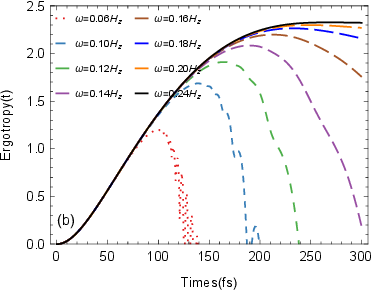}\hspace{0in}%
\end{minipage}\hspace{-0.5cm} 
\begin{minipage}{0.9\columnwidth}\centering
\includegraphics[width=0.5\columnwidth] {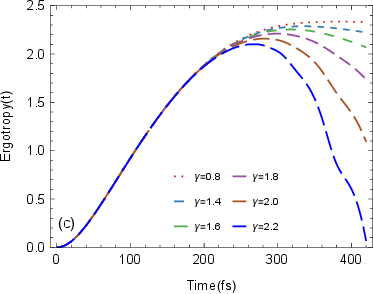}\hspace{0in}\includegraphics[width=0.5\columnwidth] {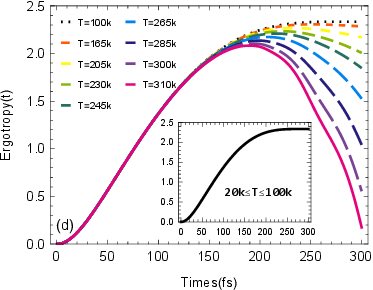}\hspace{0in}%
\caption{Ergotropy dynamics under different environmental parameters: (a) cutoff frequency $\omega_0$, (b) spectral profile influence via Redfield tensor, (c) coupling strength $\gamma$, and (d) environmental temperature $T$.}\label{Fig4}
\end{minipage}
\end{figure}

While the preceding section elucidated the role of intrinsic system parameters and external driving in modulating the charging performance of many-body quantum batteries (QBs), it is equally essential to investigate how environmental factors affect the ergotropy dynamics. This is especially relevant in realistic scenarios where the QB system is inevitably open and interacts with its surrounding environment. In this section, we analyze how different environmental control parameters-including the spectral cutoff frequency $\omega_0$, system-environment coupling strength $\gamma$, environmental spectral profile $J(\omega_k)$, and thermal temperature $T$-influence the ergotropy evolution, as shown in Fig.~\ref{Fig4}.

\paragraph*{(a) Cutoff Frequency $\omega_0$:}~The environmental spectral density is modeled via a Debye-type function, characterized by a cutoff frequency $\omega_0$ that defines the spectral bandwidth of environmental fluctuations. Fig.~\ref{Fig4}(a) illustrates that increasing $\omega_0$ effectively enhances the system¡¯s capability to reach a steady ergotropy level. This behavior can be attributed to the redistribution of environmental modes: higher $\omega_0$ values provide a broader spectrum of modes capable of interacting with the system, thereby promoting decoherence processes that suppress residual oscillations in the ergotropy. However, this suppression is not purely detrimental¡ªin the presence of strong many-body coherence, moderate decoherence can help the system relax to a quasi-steady energetic configuration, improving the effective work storage.

The cutoff frequency $\omega_0$ in the Debye spectral density $J(\omega_k) = \gamma \frac{\omega_k}{\omega_0} e^{-\omega_k/\omega_0}$ determines the range of environmental modes that can couple to the QB system. As shown in Fig.~\ref{Fig4}(a), increasing $\omega_0$ results in more stable ergotropy dynamics, reducing oscillations and accelerating convergence to a steady value. Physically, this suggests that a broader environmental spectrum facilitates rapid decoherence of non-energy-contributing coherences, thereby stabilizing the useful work component. However, this stabilization does not imply performance enhancement: the peak ergotropy is slightly suppressed, indicating a trade-off between stability and maximal extractable work.

\paragraph*{(b) Redfield Tensor Effect and Spectral Frequency:}~Fig.~\ref{Fig4}(b) reveals an apparent contradiction with Fig.~\ref{Fig2}, where increasing $\omega_0$ was associated with a weakening of the Redfield dissipator, suggesting a reduction in dissipation-induced decoherence. This apparent discrepancy arises due to a subtle interplay: while a weakened Redfield tensor suggests less energy leakage from the system, the broader environmental spectrum associated with large $\omega_0$ simultaneously accelerates decoherence of fast-varying system coherences. This dichotomy underlines the importance of matching the environmental timescale with the intrinsic dynamical timescale of the QB. Thus, environmental engineering via $\omega_0$ provides a flexible knob to either stabilize or destabilize ergotropy, depending on the system¡¯s operating regime.

Fig.~\ref{Fig4}(b) complements the interpretation by revealing how $\omega_0$ modifies the structure of the Redfield dissipator. Increasing $\omega_0$ reduces the dissipative contribution from the Redfield tensor, but still results in smoother ergotropy profiles. This reveals a key physical insight: while dissipation weakens, decoherence still operates effectively through environmental dephasing, highlighting that ergotropy stabilization in this regime is not necessarily driven by stronger dissipation, but rather by more efficient phase randomization across system modes.

\paragraph*{(c) System-Environment Coupling Strength $\gamma$:}~From the analytical form of the spectral density $J(\omega_k) = \gamma \frac{\omega_k}{\omega_0} e^{-\omega_k/\omega_0}$, the coupling constant $\gamma$ directly scales the strength of interaction between the many-body QB and the bath. Fig.~\ref{Fig4}(c) demonstrates that increasing $\gamma$ leads to a pronounced degradation in the peak ergotropy and accelerates its decay. Physically, stronger coupling intensifies energy backflow from the system to the environment and enhances dissipative processes that erode the coherence and population inversion necessary for storing usable energy. This result signifies that even if the environmental spectral profile is engineered favorably, excessive coupling nullifies potential gains by introducing irreversible losses.

The system-environment coupling constant $\gamma$ enters linearly into $J(\omega_k)$ and controls the intensity of interaction between the many-body QB and its surroundings. As evident in Fig.~\ref{Fig4}(c), increasing $\gamma$ leads to a substantial drop in both the peak and steady-state ergotropy. Stronger coupling amplifies dissipative energy leakage and accelerates decoherence, rapidly eroding quantum correlations that contribute to usable energy. This effect is particularly severe in many-body QBs, where long-range entanglement or coherent excitations across units are highly susceptible to environmental noise. Thus, $\gamma$ acts as a destructive factor when exceeding a critical threshold, pointing to the need for minimal, precisely engineered coupling in practical device design.

\paragraph*{(d) Environmental Temperature $T$:}~Finally, Fig.~\ref{Fig4}(d) exhibits the temperature-dependent behavior of the QB system. At cryogenic temperatures ($T \lesssim 100$~K), the evolution of ergotropy remains nearly unaffected, indicating a thermally protected regime dominated by quantum coherent dynamics. However, as the temperature exceeds this threshold ($T \gtrsim 100$~K), the system exhibits severe deterioration in energy storage capacity. This thermal decoherence stems from enhanced phonon occupation at higher $T$, which couples to system transitions and scrambles the quantum correlations responsible for ergotropy. These results underscore the necessity of low-temperature operation for preserving quantum advantages in battery performance, especially when operating in the strong-interaction or coherence-assisted charging regimes.

Thermal noise from a bosonic bath introduces additional decoherence pathways. Fig.~\ref{Fig4}(d) shows that below $T \approx 100\,\mathrm{K}$, the influence of temperature on the ergotropy dynamics is negligible, indicating the dominance of quantum-coherent evolution in this regime. However, as $T$ increases beyond this threshold, the extractable work rapidly diminishes. Elevated temperatures increase thermal occupation of environmental modes, thereby enhancing incoherent transitions and washing out population inversion and coherence in the QB. This underscores that thermal stability is a critical condition for operating quantum batteries in the high-performance regime. Low-temperature environments not only suppress thermally induced decoherence but also preserve the coherence-assisted charging effects characteristic of many-body QBs.

The results in Fig.~\ref{Fig4} collectively indicate that environmental factors exert a dual-edged influence on QB performance. While certain environmental modes (e.g., broad $\omega_0$) can help stabilize the output, excessive system-environment interaction ($\gamma$) and thermal agitation (high $T$) are detrimental to energy retention. These insights suggest that optimal ergotropy retrieval in many-body QBs necessitates a fine balance: designing environments that are spectrally rich yet weakly coupled and thermally isolated.

The above findings underscore the dual role of environment in QB operation. On one hand, an appropriately structured environment-e.g., a broad but weakly coupled spectrum¡ªcan promote steady energy output by selectively damping non-useful dynamics. On the other hand, excessive coupling strength or thermal noise can rapidly suppress ergotropy. The insight that $\omega_0$ and $T$ play opposite roles on different time scales offers design flexibility. Specifically, materials with tunable spectral responses or system-bath decoupling techniques can be leveraged to optimize energy storage while preserving ergotropy.

\subsection{ Possible experimental realization }

Quantum batteries (QBs) are emerging as energy-efficient quantum devices capable of storing and releasing energy through coherent and entangled dynamics~\cite{mondal2021manybody,Shi2022EntanglementCA,PhysRevA.107.023725}. In this work, we have introduced a many-body QB model based on a collective $\Lambda$-type three-level system, where multiple subsystems share a common excited state. To address concerns regarding the physical foundation and experimental realizability of such a model, we now detail feasible implementation strategies based on current experimental platforms.

A promising candidate for realizing our proposed $\Lambda$-type collective configuration is an array of semiconductor quantum dots. Quantum dots can be engineered to support discrete energy levels with tunable energy gaps via external gate voltages and strain fields. Importantly, exciton states in coupled quantum dots can act as a shared excited state, while two spatially localized ground states serve as the lower levels of each $\Lambda$ system~\cite{Krenner2005Direct,Stinaff2006Optical}. Coupling between dots can be engineered via tunneling barriers or dipole-dipole interactions, enabling coherent population transfer and entangled excitation dynamics.

Another well-established platform is Rydberg atomic ensembles, where individual atoms possess long-lived ground states and collectively couple to a common highly excited Rydberg state through dipole blockade mechanisms. This results in an effective $\Lambda$-type configuration with a shared excited state and strong inter-atomic correlations~\cite{Saffman2010Quantum,Comparat2010Dipole}. Recent experiments have demonstrated coherent excitation exchange and energy transport in such systems, making them a viable architecture for testing the collective ergotropy effects described in our model~\cite{Bernien2017Probing,Omran2019Generation}.

In both platforms, the essential requirements of our model can be met: (i) tunable energy gaps across identical subsystems, (ii) controllable tunneling or coupling strengths, and (iii) access to a common excited state for all subunits. These features are critical to reproducing the ergotropy dynamics and coherent charging behavior predicted in our simulations.

We also note that superconducting circuit QED platforms can simulate effective $\Lambda$ systems using transmon qubits coupled to resonators, with the shared cavity mode acting as a common excited level~\cite{Koch2007Charge,Blais2021Circuit}. Such systems provide high control fidelity and programmable interactions, and have been used to study quantum battery concepts experimentally~\cite{PhysRevA.107.023725}.

In summary, the proposed $\Lambda$-type many-body QB model is not only of theoretical interest but also grounded in experimentally accessible platforms. The collective structure can be realized using quantum dots, Rydberg atoms, or superconducting qubit-resonator circuits, each offering a viable path for exploring the quantum thermodynamic performance of collective QB architectures.

\section{Conclusion}\label{sec:conclusion}

In conclusion, we have proposed a strongly coupled many-body QB model that enables rapid and stable collective charging while effectively suppressing discharge reversals. This model incorporates a non-perturbative treatment of system-environment interactions and allows us to quantitatively analyze the dynamics of ergotropy under various driving and environmental parameters. Our findings demonstrate that enhancing the driving frequency ($\Omega$) and tunneling strength ($T_e$) can significantly accelerate and stabilize energy storage. In contrast, strong bandgap detuning ($\Delta E$), excessive driving amplitude ($V$), and increased environmental coupling ($\gamma$) may induce non-adiabatic transitions or dissipative effects that degrade performance. Importantly, our results reveal charging dynamics that deviate from classical macroscopic batteries, highlighting intrinsic quantum many-body features that emerge under strong coupling. These effects are concretely characterized through ergotropy profiles and coherence-dependent evolution timescales. Furthermore, our identification of parameters that regulate or suppress energy reversal provides valuable guidance for experimental realization.

Looking ahead, this work suggests promising directions for employing advanced non-perturbative methods, such as tensor-network techniques or HEOM approaches-and motivates future efforts to implement and test these findings in solid-state, photonic, or superconducting platforms. Combined with the proposed experimental schemes, our model may assist the design of programmable QB schemes in experimentally accessible settings.

\section*{Author contributions}
S. C. Zhao conceived the idea. Y. F. Yang performed the numerical computations and wrote the draft, and S. C. Zhao did the analysis and revised the paper. N. Y. Zhuang participated in part of the discussion.
\section*{Data Availability Statement}

This manuscript has associated data in a data repository. [Authors' comment: All data included in this manuscript are available upon resonable request by contaicting with the corresponding author.]

\section*{Conflict of Interest}

The authors declare that they have no conflict of interest. This article does not contain any studies with human participants or animals performed by any of the authors. Informed consent was obtained from all individual participants included in the study.

\begin{acknowledgments}
We gratefully acknowledge support of the National Natural Science Foundation of China ( Grant Nos. 62065009 and 61565008).
\end{acknowledgments}





\bibliography{reference}
\bibliographystyle{apsrev4-1}

\end{document}